\documentclass{ws-ijmpd}
\usepackage{epsfig}

\usepackage{epsfig}
\usepackage{amssymb}

\begin{document}
\markboth{Tina Kahniashvili}{Gravitational Radiation  from Primordial
Helical Turbulence}
\title{Gravitational Radiation  from Primordial
Helical Turbulence}

\author{Tina Kahniashvili}
\address{Department of Physics, Kansas State University,
116 Cardwell Hall, Manhattan, KS 66506, USA
\\ Center for Plasma Astrophysics, Abastumani Astrophysical
Observatory, 2A Kazbegi Ave, GE-0160 Tbilisi, Georgia}

\date{September 2005 \hspace{0.3truecm} KSUPT-05/4}
\maketitle
\begin{abstract}
In present talk I show that  primordial helical turbulence produced
 during a first-order phase transition
induces circularly polarized cosmological gravitational waves (GWs).
The degree of
 polarization as well as the characteristic frequency and the amplitude
of these GWs depend crucially on phase-transition model
of primordial turbulence. I present a brief discussion on the possibility of detection.

\end{abstract}

\section{Introduction}
There are several astrophysical observations indicating that the
magnetic field in the Sun, and some galaxies and  clusters of galaxies 
 might have an
helical structure \cite{astro}. On the other hand, recently  it
has been realized that magnetic (or/and kinetic) helicity can be
generated during an  early epoch of the universe 
\cite{helicity,jedamzik}. Primordial helical motions influence
magneto-hydrodynamical processes in the early plasma as well as
cosmological perturbation dynamics \cite{helicity2,dynamo}.
 Since cosmological gravitational waves (GWs) propagate without significant
interaction after they are produced, once detected they should provide
 a powerful tool for studying the early
Universe at the time  of GW generation  \cite{tasi}. For the case at hand,
GWs are considered as a test for primordial helical turbulence.
The  present formalism is
general and can be applied to study the generation of stochastic
GWs  by
any helical vector field (e.g.,
helical magnetic fields \cite{pvw02,cdk04,kr05}).

It should be noted that various mechanisms  for GW generation have been
studied,
including: quantum fluctuations
during inflation \cite{inflation};
  bubble wall motion
and collisions during phase transitions \cite{kos};
 cosmological magnetic fields \cite{magnet,kmk02};
 and plasma turbulence
\cite{kmk02,dolgov,nic}.
 Here I focus on gravitational radiation generated by primordial helical
turbulence  \cite{helicity,kr05}. This talk is based on 
results obtained in collaboration with G. Gogoberidze, A.
Kosowsky, A. Mack, and B. Ratra (see Refs.~\cite{kmk02,kgr05}). We
find that helical turbulence generates circularly polarized
stochastic gravitational waves (GWs) and we compute the
polarization degree.
 Primordial polarized  GWs
might be  generated from quantum fluctuations accounting for
 the gravitational
Chern-Simons term \cite{rodriguez}.

As it is well known GWs are generated by the transverse and
traceless part of the stress-energy tensor $T_{\mu \nu}$
\cite{mtw73}. In our case $T_{\mu \nu}$ describes  a turbulent
cosmological fluid after a
 phase transition \cite{kos,kmk02,dolgov,nic}.
 For spatial indices $i \neq
j$, $T_{ij}({\bf x}) = (p+\rho) u_i({\bf x}) u_j({\bf x})$,  where $p$ and
$\rho$ are the fluid pressure and  energy density and
 ${\mathbf u}({\mathbf x})$
 is the fluid velocity.
 The fluid enthalpy density $p+\rho$ is taken  to be
constant throughout space.
The transverse and traceless part of
$T_{ij}$ in Fourier space is  $ \Pi_{ij}({\mathbf
{k}},t)=[ P_{il}({\mathbf{\hat k}})P_{jm}({\mathbf{\hat k}})
-\frac{1}{2} P_{ij}({\mathbf{\hat k}}) P_{lm}({\mathbf{\hat k}})
]T_{lm}({\mathbf{ k}},t)$, where
 the projector onto the transverse
plane $P_{ij}({\mathbf{\hat k}}) = \delta_{ij}-{\hat k}_i {\hat
k}_j$ with  ${\hat k}_i = k_i/k$, where $k_i$ is the wave-vector.
Consistent with observations we have assumed flat space sections.

Since turbulence has a stochastic nature the induced GWs should be
also stochastic. They will form relic GWs background, and if there is
a parity breaking in the source, one can expect the circular polarization of
such waves \cite{kgr05}.
 To compute the gravitational waves signal  from helical turbulence,
first the model of turbulence should be specified.

\section{Model helical turbulence}

To model the turbulence we assume  that in the early Universe at time
$t_{\rm{in}}$ (at a phase transition)
liberated vacuum energy $\rho_{\rm{vac}}$
is converted into  (turbulent) kinetic
energy of the cosmological plasma  with an efficiency
$\kappa$ over a time scale $\tau_{\rm{stir}}$
 on a  characteristic source length scale $L_S$ 
 \cite{kos}.
 After  generation, the turbulence kinetic energy  cascades from
larger to smaller scales. The cascade stops at a damping scale,
$L_D$, where the turbulence  energy is removed by
dissipation.
As usual, we  assume
 that the turbulence  is produced in a time much less
than the Hubble time,  $\tau_{\rm{stir}} \ll 1/H_{\rm{in}}$ --- here
$H_{\rm{in}}$ is the Hubble parameter at $t_{\rm{in}}$
\cite{kmk02,dolgov}, and rapidly generates GWs. We therefore
ignore the expansion  of the Universe  when
 studying the  generation of GWs (see below).

To compute the induced GW power spectrum one must have the
 source two-point function
$\langle \Pi_{ij}^\star ({\mathbf k}, t) \Pi_{lm}
({\mathbf k^\prime},t^\prime )\rangle $.  This is determined
by the fluid velocity two-point  correlation function.
For stationary,
isotropic and homogeneous flow the velocity
two-point  function is \cite{pvw02,kr05}
\begin{equation}
\langle u^\star_i({\mathbf k})u_j({\mathbf k'})\rangle ={(2\pi)^3}
\delta^{(3)}({\mathbf k}-{\mathbf k'}) [P_{ij} P_S(k) + i \epsilon_{ijl}
\hat{k}_l P_H(k)]. \label{spectrum}
\end{equation}
Here $P_S(k)$ and $P_H(k)$ are
the symmetric (related to the
kinetic energy density per unit enthalpy of the fluid)
 and helical (related to the  average
kinetic helicity $\langle \mathbf{u}
 \cdot (\nabla \times \mathbf{u}) \rangle$) parts of the
 velocity
power spectrum \cite{pvw02,kr05}. Causality requires $P_S(k)
\geq |P_H(k)|$ (Schwartz inequality, see  p.~161 of Ref.~\cite{L}, and
Refs.~\cite{k73,MC96,helicity,jedamzik,helicity2}).

However, in the present case,  the source of turbulence
 acts for only a short time $\tau$, possibly not  exceeding
the large-scale eddy turnover time $\tau_S$ (corresponding  to
 length scale  $L_S$) --- for self-consistency, however,
 we assume that the source is active
 over a time $\tau =  \mbox{max}( \tau_{\rm{stir}}, \tau_S)$
 \cite{kmk02} --- resulting in a time-dependent velocity spectrum.
To model the development  of  helical
turbulence during the time interval
 $(t_{\rm{in}}, t_{\rm{fi}}=t_{\rm{in}}+\tau)$
we make several simplifying assumptions:

(a) Turbulent fluid kinetic energy is present on all
scales in the inertial (Kolmogorov)
 range  $k_S<k<k_D$. Here $k_S=2\pi/L_S$ and $k_D=2\pi/L_D$,  and
the inertial range includes
length scales $L \in (L_D, L_S)$, i.e., length scales shorter than those
on which energy flows into the turbulence and larger than those on which the
turbulence energy is dissipated.
We also assume that the energy is injected into the turbulence
continuously over a time $\tau$,
rather than as an instantaneous impulse \cite{kmk02,dolgov,nic,kgr05}.

(b) Unequal time correlations are  modeled as \cite{kmk02,kgr05}
\begin{equation}
\langle u^\star_i({\mathbf k}, t) u_j({\mathbf k'}, t')\rangle
={(2\pi)^3} \delta^{(3)}({\mathbf k}-{\mathbf k'}) [P_{ij} F_S(k,
t-t^\prime)
+ i
\epsilon_{ijl} \hat{k}_l F_H(k, t-t^\prime)], \label{spectrumtime}
\end{equation}
where  the $t-t^\prime $
 dependence of the  functions $F_S$ and $F_H$
 reflects the assumption of time translation invariance.
Since energy is injected continuously, at  $t=t^\prime\in (t_{\rm{in}},
t_{\rm{fi}})$,  $F_{S}(k,0)=P_{S}(k)$ and $ F_{H}(k,0)=P_{H}(k)$.

(c) The decay of non-helical   turbulence is determined
 by a monotonically decreasing function $D_1(t)$
and $ F_S(k,t) = P_S(k) D_1(t)$, p.~259 of Ref.~\cite{H}. Extending
this assumption to the  helical turbulence case we also model
$F_{H}(k,t)=P_{H}(k) D_{2}(t)$, where
 $D_{2}(t)$ is another monotonically decreasing function.
 Since in the considered model most of the power
is
 in the inertial range \cite{kgr05},
 for simplicity we discard power outside the inertial range by truncating
 $P_S$ and $P_H$ at  $k<k_S$ and $k>k_D$, i.e., we assume
  $P_{S}$ and  $P_{H}$ (and so $F_S$ and $F_H$)
 vanish outside the inertial range.

(d) We model the power spectra by  power laws,
$P_{S}(k) \propto k^{n_{S}}$ and $P_{H}(k) \propto k^{n_{H}}$. For
 non-helical hydrodynamical turbulence  the Kolmogorov spectrum
 has $n_S=-11/3$.
 It has been speculated that in a magnetized medium   an
 Iroshnikov-Kraichnan spectrum with $n_S=-7/2$ might develop instead.
 The presence of hydrodynamical helicity makes the situation more
 complex. Two possibilities have been discussed. First,
with a
forward cascade (from large  to small scales)
 of both energy and helicity
 (dominated by energy dissipation  on small scales) one has spectral
indices $n_S=-11/3$ and $n_H=-14/3$ (the helical Kolmogorov (HK)
spectrum), p.~243 of Ref.~\cite{L}.  Second, if helicity
 transfer and small-scale helicity dissipation dominate,
 $n_S=n_H=-13/3$ (the helicity transfer (HT) spectrum) \cite{MC96}.
 The HK spectrum has been observed in the
inertial range of weakly helical turbulence
 (i.e., where $|P_H(k)| \ll P_S(k)$) \cite{BO97}.
 For strongly helical hydrodynamical turbulence
the characteristic small-scale length scale of helicity dissipation
 is larger than the
Kolmogorov energy dissipation length scale \cite{DG01}.
 Therefore here the inertial range is taken to consist of two
sub-intervals, both with power-law spectra. For  smaller
$k$  the spectra are determined by helicity
 transfer and have
  $n_S=n_H=-13/3$, while for larger $k$
turbulence becomes non-helical and the more common HK spectrum is realized.
Since GW generation is mostly determined by the physics at small $k$
(near $k_S$) \cite{kmk02,dolgov},
it is fair to only use the HT spectrum in this case also \cite{kgr05}.

Based on these considerations  we model the primordial  spectra as
$P_S(k)=S_0 k^{n_S}$ and $P_H(k)= A_0
k_S^{n_S-n_H}k^{n_H}$, where: (i) for the HK case $S_0=\pi^2C_k
{\bar\varepsilon }^{2/3}$ and $A_0=\pi^2C_k {\bar\delta}
/({\bar\varepsilon }^{1/3}k_S)$ \cite{DG01}, implying $A_0/S_0 =
{\bar \delta}/({\bar \varepsilon}k_S)$; and,   (ii) for the HT case  $S_0 =
\pi^2 C_s {\bar\delta }^{2/3}$ and $A_0 = \pi^2 C_a {\bar\delta
}^{2/3}$ \cite{MC96}. Here
 ${\bar\varepsilon}$ and  ${\bar\delta}$ are the energy and mean helicity
 dissipation rates per unit enthalpy, and $C_k$ (the Kolmogorov constant),
 $C_s$, and $C_a$ are  constants of order unity.

The formalism described here is applicable for  locally isotropic
turbulence. Another requirement is  connected with plasma
viscosity, i.e. the Reynolds number ${\mbox{Re}}$, which is related to the
cut-off scales $k_S$ and $k_D$ (for HK case,
${\mbox{Re}}=(k_D/k_S)^{4/3}$), has to be enough large
\cite{my75}. Summarizing, to model  cosmological turbulence it is
required to know \cite{kgr05}: (i) the part of the vacuum energy
converted to the turbulent motions $\kappa\rho_{\rm{vac}}$; (ii)
the ``stirring'' scale $L_S$; (iii) the temperature of the universe
$T_{\rm{in}}$ when turbulence is generated. This temperature
defines the energy density and enthalpy ($\rho+p$) of the plasma; 
 (iv) the ratio between energy and helicity dissipation 
rates ${\bar \delta}/(k_S{\bar \varepsilon})$. Knowing these 
quantities and the turbulent motion spectral index, we are able to
obtain the damping scale $k_D$ and the plasma viscosity $\nu$ (p.~483 of
Ref.~\cite{my75}). In particular,
 for locally isotropic turbulence the energy
dissipation rate  ${\bar\varepsilon} =
2\nu \int_{k_S}^{k_D} dk~k^4 P_S(k)/\pi^2$
 is equal to the source power input, i.e.,
\begin{equation}
\frac{4}{3}{\bar\varepsilon} = \frac{\kappa\rho_{\rm{vac}}}{\rho\tau},
\label{var1}
\end{equation} where $\kappa$ is the phase transition
efficiency. For HK turbulence with $n_S=-11/3$,
the damping scale is related to plasma viscosity via \cite{kmk02}
\begin{equation}
\frac{9}{2} \nu^3 \tau k_D^4= \frac{\kappa \rho_{\rm{vac}}}{\rho}
.
\label{kd1}
\end{equation}
 For the HT spectrum ($n_S=-13/3$) the  damping
scale is connected with plasma viscosity and the dissipation rates
$\bar{\varepsilon}$ and $\bar\delta$ via
$
3\nu({\bar\delta}k_D)^{2/3}=
{\bar\varepsilon}$; using the fact 
 that the total energy density of turbulent motion
$\rho_{\rm{turb}}=\kappa\rho_{\rm{vac}}$, we find \cite{kgr05}
\begin{equation}
\kappa\left(\frac{k_S}{k_D}\right)^{2/3}=3\tau \nu k_S^2~
\label{kd2}\end{equation}
The $L$-scale eddy turnover (circulation) time can be obtained as a ratio of length scale $L$ to the physical velocity $v_L$ related to the normalization of the symmetric part of turbulent
motion spectrum $P_S(k)$ (or the energy dissipation rate $\bar \varepsilon$). It is easy to find that $2\tau_L \simeq 3{\bar\varepsilon}^{-1/3}L^{2/3}$ for any $L$ inside the inertial range $L_D<L<L_S$ \cite{kmk02,kgr05}.

It should be underlined that we use the non-relativistic
turbulence model to describe processes in the early universe.
The Kolmogorov model of turbulence has been proved only for
non-relativistic velocities in plasma, and there is no standard 
 model for the relativistic case. 
We have also used  the non-compressible plasma approximation. For the
realistic case of a compressible fluid one can expect that a large
enough input of energy may lead to  the formation of shocks. On the
other hand,  our assumption that the upper limit of plasma velocity
is equal to sound speed may be justified because of significant
thermal dissipation due to  shock waves. Presuming that shock
fronts accumulate kinetic energy, the gravitational radiation
signal will be increased (relative to  the non-relativistic
case), if we account for relativistic effects \cite{kmk02,dolgov}.

\section{gravitational wave polarization degree}

Since the turbulent source acts a short time (comparing with
Hubble time-scale), the expansion of the universe can be neglected
in the GW equation of motion, then
 in wave-number space we get \cite{mtw73}
\begin{equation}
{\ddot h}_{ij}({\mathbf k}, t) + k^2 h_{ij}({\mathbf k}, t) = 16\pi
G \Pi_{ij} ({\mathbf k}, t). \label{h_evolution}
\end{equation}
Here $G$ is the Newtonian gravitational constant. We use natural
units $\hbar = 1 = c$, physical/proper wave-numbers (not co-moving
ones), and an over-dot denotes a
 derivative with respect to
time $t$.
  $h_{ij}({\mathbf k}, t)=\int d^3\!x \,
   e^{i{\mathbf k}\cdot {\mathbf x}} h_{ij}({\mathbf x},t)$
and $h_{ij}({\mathbf x}, t)=\int d^3\!k \,
   e^{-i{\mathbf k}\cdot {\mathbf x}} h_{ij}({\mathbf k},t)/(2\pi)^3$
is the Fourier transform pair of the transverse-traceless tensor metric
 perturbation which is defined in the terms of the complete
 metric perturbation
$h_{ij}=\delta g_{ij}$   subject to the conditions
  $h_{ii}=0$ and $h_{ij}{\hat
k}^j=0$. Choosing the coordinate system so that unit vector
 ${\hat {\mathbf e}}_3$ points in the
GW propagation  direction, using the usual circular polarization basis
tensors  $e^{\pm}_{ij} = -({\bf e}_1
\pm i{\bf e}_2)_i \times ({\bf e}_1 \pm i {\bf e}_2)_j/\sqrt{2}$
 \cite{mtw73},  and
defining two states
$h^+$ and $h^-$ corresponding  to right- and left-handed
 circularly  polarized
GWs,  we have $h_{ij}=h^+e^+_{ij} + h^-
e^-_{ij}$. 

Stochastic turbulent fluctuations
generate stochastic  GWs. Gaussian-distributed GWs may be characterized by
 the wavenumber-space two-point  function
\begin{equation}
\langle h^{\star}_{ij}({\mathbf k},t) h_{lm} ({\mathbf k'},t)\rangle
=(2\pi)^3 \delta^{(3)}({\bf k}-{\bf k'}) \left[ {\mathcal
M}_{ijlm} H(k,t)
+ i{\mathcal A}_{ijlm} {\mathcal H} (k,t) \right].~
\label{gw1}
\end{equation}
Here $H ({k}, t)$ and ${\mathcal H}({k}, t)$ characterize
the GW amplitude and polarization,
 $4 {\mathcal M}_{ijlm} ({\mathbf{\hat k}}) \equiv
P_{il}P_{jm}+P_{im}P_{jl}-P_{ij}P_{lm}$, and
 $8 {\mathcal A}_{ijlm}({\mathbf{\hat k}})
\equiv {\hat {\bf k}}_q (P_{jm} \epsilon_{ilq} + P_{il}
\epsilon_{jmq} + P_{im} \epsilon_{jlq} + P_{jl} \epsilon_{imq})$
are tensors, and $\epsilon_{ijl}$ is the 3 dimentional fully
antisymmetric symbol.

  The GW degree of circular polarization  is given by \cite{meszaros}
\begin{equation}
{\mathcal P}(k,t) =  \frac {\langle h^{+ \star}({\mathbf k}, t)
h^{+}({\mathbf k'}, t) -
 h^{- \star}({\mathbf k},t) h^{-}({\mathbf k'},t) \rangle}
{\langle h^{+ \star}({\mathbf k}, t) h^{+}({\mathbf k'}, t) +
 h^{- \star}({\mathbf k},t) h^{-}({\mathbf k'},t) \rangle}
=\frac{{\mathcal H}(k, t)}{H(k, t)}.~ \label{degree}
\end{equation}

  Both  $H(k,t)$ and ${\mathcal H}(k,t)$
 are  obtained by solving Eq.~(\ref{h_evolution}), and are
related to $\Pi_{ij}({\mathbf k},t)$. For instance, an
axisymmetric stochastic vector source (non-helical turbulent
motion or any other non-helical vector field) induces unpolarized
GWs with $ |h^+({\bf k}, t)|  = |h^-({\bf k}, t)|$
\cite{magnet,kmk02,dolgov}. On the other hand, the presence of
 a helical source  (i.e.,  a source that breaks parity symmetry)
 alters this situation \cite{kgr05}.

Given a  model of the turbulence, the turbulent source
two-point function is  \cite{kgr05}
\begin{equation}
\langle\Pi^{\star}_{ij}({\mathbf k},t)\Pi_{lm}
({\mathbf k'},t+y)\rangle
=(2\pi)^3 \delta^{(3)}({\bf k}-{\bf k'})\left[ {\mathcal
M}_{ijlm} f(k,y) + i{\mathcal A}_{ijlm} g(k,y) \right],
~\label{pi1}
\end{equation}
where $ {\mathcal M}_{ijlm}$ and ${\mathcal A}_{ijlm}$ are defined below
Eq.~(\ref{gw1}). The functions $f(k, y)$ and $g(k,y)$  that
 describe the symmetric and helical
parts of the two-point function are \cite{kgr05}
\begin{eqnarray}
f(k, y)&=&
\frac{(\rho+p)^2}{256\pi^6}
\!\int\!d^3p_1\!
\int\!d^3p_2 \delta^{(3)}({\bf k}-{\bf p_1}-{\bf p_2})
\left[(1+\gamma^2)(1+\right.\nonumber\\
&+&\beta^2)D_1^2(y)P_S(p_1)P_S(p_2)
+ 4\gamma \beta D_2^2(y)\left.\right]
P_H(p_1)P_H(p_2), \label{tensor-source-sym}
\\
g(k,y) &= &\frac{(\rho+p)^2 D_1(y)D_2(y)}{128\pi^6}\int\!d^3p_1\!\int\!d^3p_2
\,
\delta^{(3)}({\bf k}-{\bf p_1}-{\bf p_2})
\left[ (1+\right.
\nonumber\\& +&\gamma^2)\beta P_S(p_1)P_H(p_2)
(1+\beta^2)\gamma P_H(p_1)P_S(p_2) \left.\right],
\label{tensor-source-hel}
\end{eqnarray}
where $\gamma={\hat {\bf k}}\cdot{\hat {\bf p}_1}$ and $\beta={\hat
{\bf k}}\cdot{\hat {\bf p}_2}$.
 The antisymmetric (parity-odd) source term $g(k, y)$ vanishes
for turbulence  without helicity.

To determine $H(k, t)$ and ${\mathcal H}(k, t)$ we
 solve Eq.~(\ref{h_evolution}) assuming that there is no GW
for times  $t<t_{\rm{in}}$, i.e., we choose  as initial conditions
$h_{ij}({\mathbf k}, t_{\rm{in}})=0={\dot h}_{ij}({\mathbf k},
t_{\rm{in}})$. To compute the induced GW power spectrum we use the
averaging technique described in detail in Sec. III.B
of Ref.~\cite{kmk02}. The main points are:
 (i) the
$\delta^{(3)}({\bf k}-{\bf k'})$ in Eqs.~(\ref{gw1}) and
(\ref{pi1})  ensure statistical isotropy of the GWs and
 allow   ${\mathbf k}'$ and
${\mathbf k}$ to be interchanged;
(ii) the statistical average can be approximated by either a time  or a
space average (this is justified for locally isotropic
turbulence where the correlation function
$\langle u_i({\mathbf x}_0, t_0)u_j({\mathbf x}_0 + {\mathbf r}, t_0 + t)\rangle$
for $t \leq \tau$ does not depend on $t_0$
for relatively small $t$,
   see Sec. 21.2 of Ref.~\cite{my75});
(iii) we choose to time average since
the Green function for Eq.~(\ref{h_evolution}) and
the source term  $\Pi_{lm} ({\mathbf k},t)$ are time dependent
 (we  then
have three time integrals for the GW two-point function,
 two from the GW Green function solutions and one from
the time averaging, $\int_{t_1}^{t_1 +T} dt/T $, where $t_1$ is
 an arbitrary  time during the time interval $\tau$ when
the source is active and the averaging time $T \leq \tau$);
 (iv) we take $Tk\gg 1$ (since $k^{-1}$ is
 of order  the light crossing time for scale $L \sim k^{-1}$ while
$T$ is of order the
$L$ scale eddy turnover time). These approximations result
in Eq.~(32) of Ref.~\cite{kmk02},
\begin{equation}
\langle h_{ij}^\star ({\mathbf k}, t_{\rm{fi}} ) h_{lm} ({\mathbf
k^\prime}, t_{\rm{fi}} )\rangle \simeq \frac{(16\pi G)^2 \tau
}{2kk^\prime} \int_{t_{\rm{in}}}^{t_{\rm{fi}}} dt~ \cos (kt)
 \langle\Pi_{ij}^\star ({\mathbf k}, t_1) \Pi_{lm} ({\mathbf k^\prime},t_1+
t )\rangle. \label{gw-spectrum}
\end{equation}
 The two-point  function on the r.h.s.~of the integral
is independent of $t_1$ and
$\langle h_{ij}^\star ({\mathbf k}, t_{\rm{fi}} ) h_{lm} ({\mathbf
k^\prime}, t_{\rm{fi}})\rangle $ is proportional to the
source duration time $\tau$,
 as  expected for locally isotropic turbulence
(for a more detailed discussion see p.~358 of Ref.~\cite{my75}).

From Eqs.~(\ref{pi1})--(\ref{gw-spectrum}) we see that
 the symmetric
 $H(k,t)$ and helical
${\mathcal H}(k, t)$ parts  of the GW two-point function
 in Eq.~(\ref{gw1}) are
integrals over $y$ of $\cos (ky)
D_1^2(y)$, $\cos (ky)D_2^2(y)$, and $\cos (ky)D_1(y)D_2(y)$.
 Both  $D_1(y)$ and $D_2(y)$
are positive monotonically-decreasing functions of $y$,
 and since $\cos(ky)$ oscillates on a time
scale shorter than the characteristic decay time for the $D(y)$'s,
the integrands  oscillate and
 $\int_{t_{\rm{in}}}^{t_{\rm{fi}}} dy \cos (ky) F_a(p,
y)F_b(|{\mathbf k}-{\mathbf p}|, y) \simeq P_a(p) P_b
(|{\mathbf k}-{\mathbf p}|)/(\sqrt{2} k)$
(where $a$ and $b$ can be $S$ or $H$).
Integrating over angles, we find at $t=t_{\rm{fi}}$ \cite{kgr05},
\begin{eqnarray}
H(k) &\simeq & A \int\!dp_1~p_1\!\int\!dp_2~p_2
{\bar \Theta}\left[(1+\gamma_p^2)(1+\beta_p^2)P_S(p_1)
P_S(p_2)+\right.\nonumber\\ &+& 
4 \gamma_p \beta_p P_H(p_1)P_H(p_2) \left.\right],~ \label{tensor-sym1_a}
\\
{\mathcal H}(k) &\simeq &2A\int\!
dp_1 ~p_1\!\int\!dp_2~p_2 {\bar \Theta}
\left[(1+\gamma_p^2)\beta_pP_S(p_1)
P_H(p_2)+ 
\right.\nonumber\\ &+& 
(1+\beta_p^2)\gamma_pP_H(p_1)P_S(p_2) \left.\right].~ \label{tensor-hel1_a}
\end{eqnarray}
Here $A= \alpha \tau/(4\pi^2 k^4)$ where $\alpha =
\sqrt{2} (p+\rho)^2 (8 \pi G)^2$,
$\gamma_p = (k^2+p_1^2-p_2^2)/(2kp_1)$, $\beta_p =
(k^2+p_2^2-p_1^2)/(2kp_2)$,  ${\bar \Theta} \equiv \theta(p_1+p_2-k)
\theta(p_1+k-p_2) \theta(p_2+k-p_1)$, and $\theta$ is the Heaviside
 step function which is zero (unity) for negative (positive) argument.
Let's note that present approximations differs from ones used
previously in Refs. \cite{kmk02,cdk04}. The difference comes from
using $\delta^{(3)}({\mathbf{k}}-{\mathbf{p}})$ function to
evaluate angle integrals and don't absorb modulus integration
$|{\mathbf p}|$. This preserves to make errors when approaching
the integration edges. In particular , since the spectra
$P(|{\mathbf{k}}-{\mathbf{p}}|)$ are defined only for $k_S <
|{\mathbf{k}}-{\mathbf{p}}| < k_D$, then the integration over
spatial angle $\gamma$ can not range from $-1$ to $1$ as it is
usually done making the convolution when the spectra are defined
in whole of $p\in(0, \infty)$.

For power-law $P_{S}(k)\propto
k^{n_{S}}$ and  $P_{H}(k)\propto k^{n_{H}}$ the integrals in
Eqs.~(\ref{tensor-sym1_a}) and (\ref{tensor-hel1_a}) can be done
analytically, but the results are complicated and do not edify.  Instead
 we compute the degree of circular polarization, Eq.~(\ref{degree}), by
evaluating the integrals numerically for different
parameter values \cite{kgr05}. Results are shown in  Fig. 1.

\begin{figure}
\includegraphics{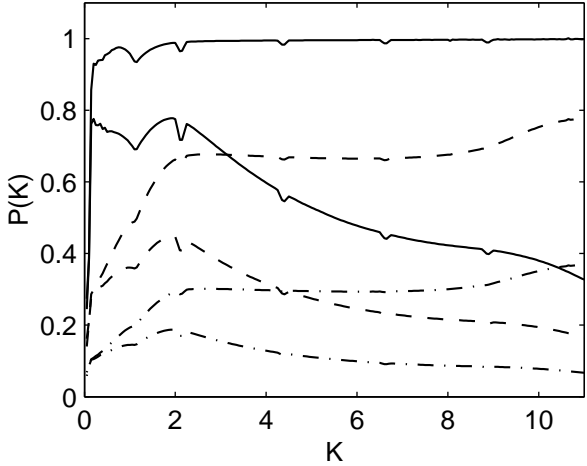}
\caption{GW polarization degree ${\mathcal P}(K, t_{\rm{fi}})$,
Eq.~(\ref{degree}), as a function of scaled wave-number $K=k/k_S$
relative to the large-scale wave-number $k_S$ on which energy is
pumped into the turbulence. This is evaluated at time
$t_{\rm{fi}}$, after the turbulence has switched off, and remains
unchanged to the present epoch. It has been computed for a damping
wave-number $k_D=10k_S$. Three pairs of curves are shown. Solid
lines correspond to the amplitude ratio $A_0/S_0=1$, dashed lines
to $A_0/S_0=0.5$, and dot-dashed lines are for $A_0/S_0 =0.2$. The
upper line in each pair corresponds to HT turbulence with
$n_S=n_H=-13/3$ and the lower line to HK turbulence with
$n_S=-11/3$ and $n_H=-14/3$. Even for helical turbulence with
$A_0/S_0 \leq 0.5$,  for large wave-numbers $k \sim k_D$,
$n_S=n_H=-13/3$ is unlikely so the large $K$ part of the lower
dashed and dot-dashed HT curves are unrealistic. The large $k \sim
k_D$ decay of the HK curves is a consequence of vanishing helicity
transfer at large $k$ }
\end{figure}

Figure 1 and other  numerical results show that for  maximal
helicity turbulence (when $A_0=S_0$) with equal spectral indices
$n_H = n_S<-3$, the polarization degree
${\mathcal P}(k) \simeq 1$ (upper solid line).
 For weaker helical turbulence (when $ A_0 <
S_0$) with $n_H \simeq  n_S<-3$,  ${\mathcal P(k)}
\rightarrow CA_0/S_0$, where $1<C(n_S,n_H)<2$ is a numerical factor that
depends on the spectral indices. For HT turbulence with
$n_S=n_H=-13/3$,  $C \approx 1.50$, while  for Iroshnikov-Kraichnan
MHD turbulence ($n_S=n_H=-7/2$), $C\approx1.39$. Excluding the edges of  the
inertial range  $k_S < k <k_D$,
an analysis of
Eqs.~(\ref{tensor-sym1_a}) and (\ref{tensor-hel1_a}) shows that the main
contribution to the integrals come from  areas with $p_1\sim
k_S,~p_2\sim k$ and $p_1\sim k,~p_2\sim k_S$.
In this case
(for arbitrary spectral indices $n_S$, $n_H<-3$)
 $H(k),~{\mathcal H(k)} \propto k^{n_S-3}
k_S^{n_S+3}$,

\section{Results and discussion}

To make the connection with the GW measurements, we have to define
the real-space two point correlation function
$\langle  |h^\pm  ({\mathbf x},t_{\rm{end}})|^2 \rangle $. We find
\begin{equation}
\langle  |h^\pm  ({\mathbf x},t_{\rm{fi}})|^2 \rangle \simeq
-\frac{ 9 \alpha \tau S_0^2 k_S^{n_S+3}}{64\pi^4 (n_S+3)}
\int_{k_S}^{k_D}dk (1 \pm {\mathcal P}(k))k^{n_S-1}.
\label{H43}
\end{equation}
GW amplitudes  are  conventionally expressed as
 $
\langle  h^{ij}  ({\mathbf x},t_{\rm{fi}})
 h^{ij}  ({\mathbf x},t_{\rm{fi}}) \rangle
  = 2 \int_0^\infty d{\rm{ln}}f [|h^{+}(f)|^2 + |h^{-}(f)|^2],
$ see Eq.~(11) of Ref.~\cite{m00}, where
the frequency $f$ of a GW generated
  by an eddy of length $L$ is
$f=\tau_L^{-1}$ where $\tau_L$ is the eddy turnover time
\cite{kos,kmk02,dolgov}.
Using $f=2{\bar\varepsilon}^{1/3} L^{-2/3}/3$ we get
\begin{equation}
f \simeq \frac{1}{2\pi^2}\sqrt{\frac{S_0
k^{n_S+5}}{|n_S+3|}}, ~~~~~~~~
f_S \simeq 2L_S^{-\frac{n_S+5}{2}}
\sqrt{\frac{ (2\pi)^{n_S+1} S_0}{|n_S+3|}},
\label{frequency}
\end{equation}
where $f_S$ is the frequency that corresponds to the stirring length $L_S$.
Both the frequency and the amplitude of GWs are  inversely  proportional to
the cosmological scale factor, so the
frequency ${\bar f}$ today and $f$  when the temperature was
$T_{\rm{in}}=100~T_{100}$ GeV are related by
${\bar f} = 1.65 \times 10^{-5}  T_{100} g_{100}^{1/6}
f/H_{\rm{in}}$ Hz  \cite{m00},
 where $g_{\rm{in}} = 100 g_{100}$ is the number of
 relativistic degrees of freedom at
$t_{\rm{in}}$. Since we truncate turbulence power for $L>L_S$,
the GW spectrum is non-zero only for
${\bar f}>{\bar f}_S$, where \cite{kgr05}
\begin{equation}
{\bar f}_S =  1.9 \times 10^{-6}
\sqrt{\frac{n_S+5}{|n_S+3|}}
\left(\frac{{\bar\varepsilon}}{ \nu}\right)^{{1}/{2}}
\left(\frac{L_D}{L_S}\right)^{(n_S+5)/{2}}
T_{100}~g_{100}^{{1}/{6}}~H_{\rm{in}}^{-1}~{\mbox{Hz}}
\label{fs}
\end{equation}
Here we used Eqs.(\ref{var1})-(\ref{kd2}). For HK turbulence
($n_S=-11/3$) the expression for ${\bar f}_S$ will be transformed
in Eq.~(53) of Ref.~\cite{kmk02}.

Using Eqs.~(\ref{H43}) and (\ref{fs}),
 and neglecting the weak $k$-dependence
of the GW polarization degree ${\mathcal P}$,  we find
 that $h^\pm ({\bar f}) \propto {\bar f}^{-11/4}$ for the HK case
 \cite{kmk02,dolgov}, while for HT turbulence  $h^\pm ({\bar f})\propto
 {\bar f}^{-13/2}$. We expect such  a steeper dependence on
frequency for helicity induced
GWs, since the helicity transfer rate is more important on
larger scales.  In both cases the amplitude of the
GW spectrum peaks at the stirring frequency ${\bar f}_S$.

To examine the prospect of such circularly polarized GWs detection, we compute
the energy density parameters of GWs. This depends strongly on the
frequency band.
The GW energy density parameter for frequency
${\bar f}$,
$\Omega_{\rm{GW}} ({\bar f})$ is given by (see Eq.~(7) of Ref~
\cite{m00})
\begin{equation}
\Omega_{\rm{GW}}({\bar f}) h^2 = 5.9 \times 10^{35}
(|h^+({\bar f})|^2 + |h^-({\bar f})|^2)(\frac{\bar f}{\mbox{Hz}})^2,
\end{equation} where
$h$ is the Hubble constant in  units of $100$ km~sec$^{-1}$ Mpc$^{-1}$.
In our case,
\begin{eqnarray}
\Omega_{\rm{GW}}({\bar f}) h^2 &\simeq
&   1.05 \times 10^{-11}~g_{100}^{-{1}/{3}}
\left(\frac{L_S^2}{\tau H_{\rm{in}}^{-1}}\cdot \frac{n_S+5}{|n_S+3|}\right)^2
\left(\frac{L_D}{L_S}\right)^{3(n_S+5)}
\left(\frac{3\kappa \rho_{\rm{vac}} L_S}{4\nu \rho}\right)^3
\nonumber\\&&
\left(\frac{\bar{f}}{\bar{f}_S}\right)^{2(2n_S+5)/(n_S+5)}~.\label{energy}
\end{eqnarray}
As it is expected
the stirring frequency ${\bar f}_S$ and the GW spectrum are very sensitive
to phase transition properties (for details see Fig.~1 of Ref.~\cite{dolgov}).
If the phase transition is strongly first order,
 ${\bar f}_S  \simeq 5 \times 10^{-3}$Hz (for the HK case) \cite{kmk02},
is near the LISA  frequency range, but the amplitude  of the GW
signal  is below  LISA sensitivity \cite{dolgov,nic,tasi}. Also
the gravitational radiation signal coming from white dwarf
binaries \cite{white-dwarf} will be compatible by amplitude with
relic signal considered above (for the frequency range around
$10^{-3}-10^{-2}$ Hz making difficult to distinguish the source of
gravitational waves. Thus
 it is unlikely that the circularly-polarized GWs considered here
will be detected in the near future, however, GWs generated by
helical turbulence will have a enough high degree of circular
polarization and future detector configurations \cite{GWdetection}
 may well be able to.

\bigskip
The author thanks the organizers of the ICTP conference
``Non-linear Cosmology: Turbulence and Fields'' for hospitality.
The author acknowledges her collaborators G. Gogoberidze, A.
Kosowsky, A. Mack, and B. Ratra, and thanks A.~Brandenburg, C.
Caprini, A. Dolgov, R. Durrer, D. Grasso, K. Jedamzik, T.
Vachaspati, and L. Weaver for discussions. This work is supported
by CRDF-GRDF grant 3316, NSF CAREER grant AST-9875031, and DOE
EPSCoR grant DE-FG02-00ER45824.

\end{document}